\documentclass[conference]{IEEEtran}
\IEEEoverridecommandlockouts
\usepackage{cite}
\usepackage{amsmath,amssymb,amsfonts}
\usepackage{algorithmic}
\usepackage{textcomp}
\usepackage{subfig}
\usepackage{subfloat}
\usepackage{graphicx}
\usepackage[dvipsnames]{xcolor}
\usepackage{tabularx}
\usepackage{multirow}
\usepackage{booktabs} 
\usepackage{pifont} 
\newcommand{\sys}{REMARK-LLM}
\newcommand{\baseend}{AWT}
\newcommand{\baseinfer}{KGW}
\newcommand{\baserule}{CATER}
\newcommand{\baseinferrb}{EXP}
\newcommand{\cmark}{\ding{51}}%
\newcommand{\xmark}{\ding{55}}%
\def\BibTeX{{\rm B\kern-.05em{\sc i\kern-.025em b}\kern-.08em
    T\kern-.1667em\lower.7ex\hbox{E}\kern-.125emX}}
\begin{document}

\title{Watermarking Large Language Models and the Generated Content: Opportunities and Challenges}

\author{\IEEEauthorblockN{Ruisi Zhang}
\IEEEauthorblockA{University of California, San Diego \\
ruz032@ucsd.edu
}
 \and
\IEEEauthorblockN{Farinaz Koushanfar}
\IEEEauthorblockA{University of California, San Diego \\
farinaz@ucsd.edu}}

\maketitle

\begin{abstract}
The widely adopted and powerful generative large language models (LLMs) have raised concerns about intellectual property rights violations and the spread of machine-generated misinformation. Watermarking serves as a promising approch to establish ownership, prevent unauthorized use, and trace the origins of LLM-generated content. This paper summarizes and shares the challenges and opportunities we found when watermarking LLMs. We begin by introducing techniques for watermarking LLMs themselves under different threat models and scenarios. Next, we investigate watermarking methods designed for the content generated by LLMs, assessing their effectiveness and resilience against various attacks. We also highlight the importance of watermarking domain-specific models and data, such as those used in code generation, chip design, and medical applications. Furthermore, we explore methods like hardware acceleration to improve the efficiency of the watermarking process. Finally, we discuss the limitations of current approaches and outline future research directions for the responsible use and protection of these generative AI tools.
\end{abstract}

 
\section{Introduction}
In recent years, the rapid development and widespread adoption of generative large language models (LLMs) have revolutionized various domains, including natural language processing~\cite{touvron2023llama,abdin2024phi}, content creation~\cite{zhou2024survey,shen2024data}, and code generation~\cite{roziere2023code,luo2023wizardcoder}. These powerful AI tools have demonstrated remarkable abilities in generating coherent, contextually relevant, and human-like text, leading to their application in diverse fields such as chip design~\cite{nakkab2024rome,liu2023chipnemo,zhang2024automated}, medical research~\cite{wang2024jmlr,panagoulias2024evaluating,yuan2024continued}, and software development~\cite{lin2024llm,xia2024agentless}. However, the ease of access to and use of these models has raised significant concerns regarding the potential violation of intellectual property rights and the spread of machine-generated misinformation.

While commercial LLMs are typically deployed via cloud APIs~\cite{schulman2022chatgpt}, providing users with only black-box access, there are scenarios where models are open-source~\cite{OpenOrca,touvron2023llama,abdin2024phi} or distributed and embedded in edge devices~\cite{lin2024qserve,lin2024awq}. In these contexts, monitoring the distribution of LLM models, particularly open-source and embedded models, has become critical for several reasons. Firstly, model owners need to safeguard their intellectual property and ensure appropriate attribution\cite{sheybani2023zkrownn,chen2019deepattest,zhang2024emmark}, given the significant investments in research, development, and computational resources~. Furthermore, tracking model distribution facilitates effective version control and helps prevent violations of open-source licensing agreements. In embedded model deployment~\cite{aldahmani2023cyber,caviglione2023emerging}, distribution monitoring is crucial for mitigating unauthorized modifications or malicious tampering that could potentially compromise model performance or introduce detrimental behaviors. This is essential in maintaining the integrity and intended functionality of LLMs across diverse deployment scenarios.

Monitoring the distribution of LLM-generated content data is equally critical when the LLMs are served as cloud APIs~\cite{kirchenbauer2023watermark,zhang2023remark,huo2024token}. As LLMs become more sophisticated in generating human-like text across various domains, there is an increasing risk of AI-generated misinformation~\cite{chen2023can,chen2023combating}, deepfakes~\cite{vp2024llm,cai2023av}, and other forms of synthetic content being widely spread. Tracking the origin and distribution of such content is essential for maintaining the integrity of information ecosystems and protecting against potential manipulation or fraud. In domain-specific applications, such as code generation, medical data, or chip design, monitoring LLM-generated data becomes even more crucial due to the additional engineering effort in data collection~\cite{jiang2024circuitnet,zeng2020meddialog} and increasing reasoning capacity~\cite{liu2023reason,nam2024using}. Implementing robust monitoring mechanisms for LLM-generated content and data will increase the trustworthiness of the AI systems, facilitate accountability, and enable timely interventions when misuse or errors are detected.

An ideal text watermarking framework should adhere to the following criteria~\cite{zhang2023remark}: 

\begin{itemize}
    \item \textbf{Criteria 1 Effectiveness}: The inserted watermark signatures can be seamlessly extracted.
    \item  \textbf{Criteria 2 Fidelity}: The watermarked content quality shall not be compromised. This entails that signature insertion not only preserves the original semantics but also ensures that the text coherence and consistency remain undistorted.
    \item  \textbf{Criteria 3 Efficiency}: The watermark insertion and extraction are efficient. It includes both minimal computation and time overheads to ensure  IP insertion/verification without excessive computational resources.
    \item  \textbf{Criteria 4 Robustness}: Resilience against potential threats is crucial to help LLM proprietors verify IP and trace data sources. Therefore, the signatures shall remain extractable under detection and removal attacks.
    \item \textbf{Criteria 5 Undetectability}: The watermarks are invisible upon inspection. As a result, the adversary cannot detect whether a given text or model is watermarked. 
\end{itemize}

Designing the watermarking framework for LLMs also needs to consider the efficiency and scalability of both watermark insertion and detection processes to ensure wide applicability~\cite{zhang2023systemization}. As LLMs grow in size and application~\cite{wang2024towards,li2024llm}, the computational demands of watermarking operations become increasingly significant. The software-hardware co-design approach offers promising solutions to this challenge. By optimizing watermarking algorithms for hardware acceleration techniques~\cite{zeng2024flightllm,huang2024edgellm,lai2024lcm}, such as those provided by FPGAs, GPUs or specialized AI accelerators, we can substantially improve the speed and energy efficiency of watermarking. This co-design strategy involves developing parallelizable watermarking methods while creating custom hardware architectures tailored to these tasks. It enhances performance, ensures scalability for real-time applications and large-scale deployments, and minimizes computational overhead. 

\textbf{Our Contributions.}
Our work offers a comprehensive overview of LLM watermarking, including both model and content watermarking, as in Figure~\ref{fig:overview}. We survey diverse watermarking algorithms for LLMs under various threat model settings and application scenarios. Our analysis extends to domain-specific applications, including code generation, chip design, and medical data. To address scalability, we investigate hardware acceleration strategies for efficient watermarking and verification. Finally, we assess current limitations and propose future research directions, aiming to guide the development of robust and ethical watermarking solutions that align with the rapid progress in generative AI.

\begin{figure}[!ht]
    \centering
    \includegraphics[width=0.9\columnwidth]{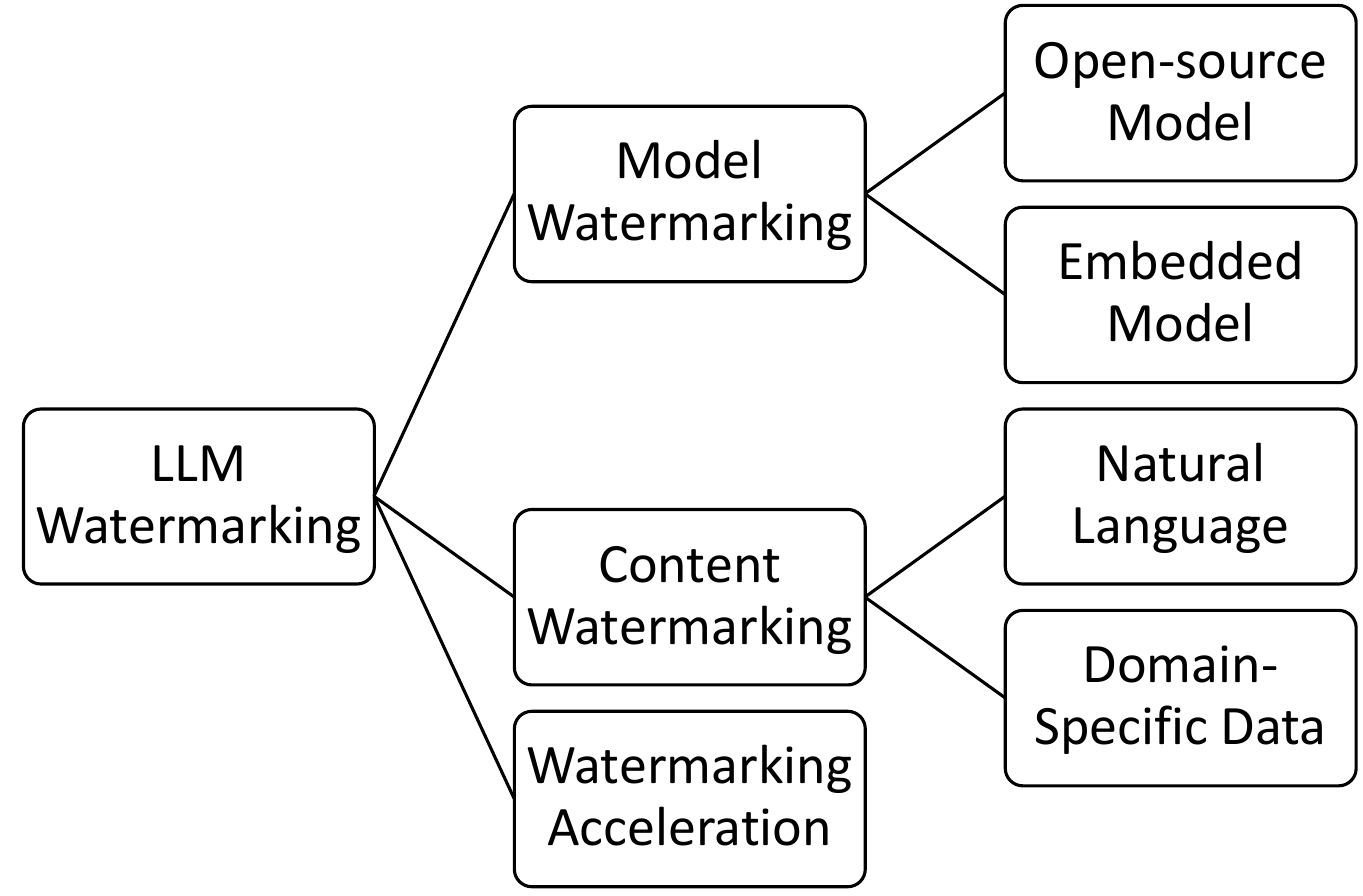}
    \caption{Paper overview. The paper first introduces LLM model watermarking and LLM-generated content watermarking. Then, it presents methods for watermark insertion and verification acceleration. }
    \label{fig:overview}
\end{figure}


\section{Watermarking Large Language Models}
In this section, we introduce related work on watermarking open-sourced LLMs and the embedded LLMs. 
\begin{figure*}[!ht]
    \centering
    \includegraphics[width= 0.95\linewidth]{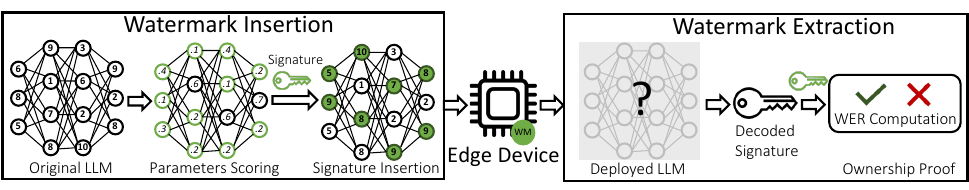}
 
    \caption{Embeded LLM watermarking overview. The watermark insertion encodes signatures into the original LLM before deployment. The watermark extraction decodes the signatures from the deployed LLM and proves ownership.}
    \label{fig:overview}
 
\end{figure*}

\subsection{Watermarking Open-source Models}
Watermarking open-source Large Language Models (LLMs)~\cite{OpenOrca,touvron2023llama,abdin2024phi} has become increasingly important as these models gain widespread adoption and distribution. The primary motivation for watermarking open-source LLMs stems from the model owners' need to ensure proper distribution and trace model usage. As these powerful language models become freely available, there's a growing concern about potential misuse, unauthorized modifications, or redistribution without proper attribution. Watermarking provides a mechanism for model owners to embed unique identifiers or traceable information within the model, allowing them to verify the authenticity of deployed instances and track how their models are being used across various applications. Moreover, watermarking can aid in detecting and preventing malicious alterations to the model, ensuring that users are interacting with the genuine, uncompromised version of the open-source LLM.

\noindent\textbf{Post-training Alignment for Watermarking}~\cite{li2024double,wei2024proving,li2023watermarking}. This method fine-tunes the pre-trained model to respond to specific prompt instructions with predetermined watermark information. The model can be aligned to generate owner information, version details, or usage terms when given a particular sequence of input tokens or a specially crafted prompt. The alignment is achieved by fine-tuning the model on a small-scale curated dataset of watermark-related question-answer pairs. The advantage is that those methodologies do not require modifying the core architecture or retraining the entire model from scratch, making it a relatively efficient and flexible watermarking solution. Additionally, this method allows for the creation of dynamic watermarks that can be updated or changed over time without extensive computation resource requirements. 

\noindent\textbf{Post-training Alignment for Fingerprinting}
~\cite{li2023plmmark,gu2022watermarking,xu2024instructional}
As a specialized subset of watermarking techniques, fingerprinting extends watermarking by aligning models with different objectives. It helps to identify different model users and fine-grained distribution of the open-source LLMs. The fingerprinting process is similar to watermarking but with a distinction in the training dataset. Rather than using a universal set of watermark-related question-answer pairs, fingerprinting employs model-specific training sets. Each model is fine-tuned on a unique, small-scale curated dataset of fingerprint-related question-answer pairs, resulting in subtle but detectable variations in output across different fingerprinted model instances.

\subsection{Watermarking Embedded Models}
Deploying generative large language models in the edge fuels the broader applications for mobile users and IoT devices. For cloud LLM APIs like ChatGPT and GPT-4, users only have black-box access to the LLMs. However, end-users on edge devices have full access to the LLMs locally. The switch in the security model leads to new threats to the LLM copyrights~\cite{zhang2024emmark}, as in Figure~\ref{fig:overview}.

The embedded LLMs deployed in the edge devices are compressed for better hardware efficiency. The most significant aspect of the compressed model from the watermarking perspective is quantization, which maps the full-precision weights into INT8/INT4 for reduced memory size and bandwidth. The sparse weight distributions introduced new challenges for model watermarking compared with full-precision models. Besides, the model pruning process may cause the model to lose the watermark that is instructed in the post-training alignment process.


\noindent\textbf{Model Weight Watermarking}~\cite{chen2020specmark,zhang2024emmark} It inserts signatures onto the trained model weights. By analyzing the model's weight distributions, the watermarking algorithm identifies the critical regions that contribute most significantly to the model's performance and output quality. Then, the watermarking insertion selectively encodes information by subtly altering the weight values in these regions. It ensures the watermark is integrated into the model's core functionality while minimizing potential degradation of overall performance. 

\noindent\textbf{Trusted Execution Environment}~\cite{wu2024secgpt,yang2024first,huang2024fast}
It offers another promising approach for protecting the intellectual property of embedded models, including large language models (LLMs). TEEs provide a secure enclave within a processor~\cite{sabt2015trusted,jauernig2020trusted,ekberg2013trusted} where sensitive computations can be performed in isolation from the rest of the system, ensuring confidentiality and integrity of the code and data within. By deploying LLMs within TEEs, model owners can enhance the protection of their intellectual property against unauthorized access or tampering. However, the deployment of LLMs in TEEs presents unique challenges due to the models' substantial size and computational requirements.  

\subsection{Potential Threats}

The adversary has full access to the watermarked LLM parameters and has knowledge of watermark insertion algorithms. However, he/she cannot access the original non-watermarked LLMs. The adversary also does not know the owners' signatures or random seeds for watermark parameter selections or the watermark dataset.

Potential threats to the watermarked LLMs include~\cite{zhang2024emmark,boenisch2021systematic,darvish2019deepsigns}:

\begin{itemize}
    \item \textit{\textbf{Attack 1 Parameter Overwriting Attacks}}: Other values replace model parameters to remove encoded watermark. 
    \item  \textit{\textbf{Attack 2 Re-watermarking Attacks}}: The adversary corrupts the original watermark by embedding new signatures.
    \item  \textit{\textbf{Attack 3 Forging Attacks}}: The adversary counterfeits fake watermarks from the watermarked model and claims the model belongs to him.
    \item  \textit{\textbf{Attack 4 Fine-tuning Attacks}}: The adversary fine-tunes the model, aiming to remove the encoded signatures.
    \item \textit{\textbf{Attack 5 Model Pruning Attacks}}: The adversary prunes or compresses the model parameters, aiming to remove the encoded watermarks.
\end{itemize}

\section{Watermarking LLM-Generated Contents}
\begin{figure*}[!ht]
    \centering
    \includegraphics[width= 0.95\linewidth]{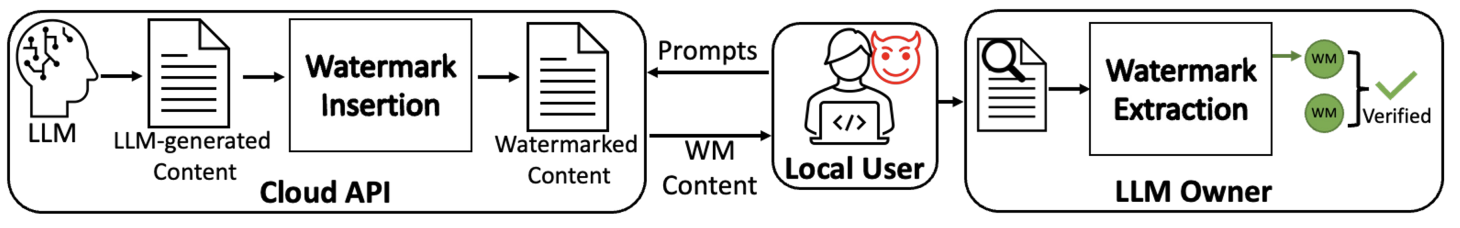}
    \caption{LLM-generated text watermarking scenario. The local user sends prompts to the remote LLM cloud API, and the API watermarks (WM) the responded texts before sending them back to users. LLM proprietor claims ownership by using the message decoding module to decode the signatures and compare them with inserted watermarks. }
    \label{fig:scenario}
 
\end{figure*}

In this section, we introduce previous work on watermarking LLM-generated natural language and domain-specific data. A high-level overview of the watermarking scenario is in Figure~\ref{fig:scenario}.

\subsection{Natural Language Data Watermarking}

Given the increasing popularity of LLMs and the heightened prominence of machine-generated media, the dangers associated with the content they produce have become more significant. Content generated by LLMs may inadvertently be employed to create counterfeit essays or inundate the internet with spam responses, thereby posing a threat to the credibility of online content.

The watermarking system~\cite{huo2024token,zhang2023remark} has wide applications with the emerging popularity of large language models: (i) Education~\cite{lancaster2023artificial}: Watermarking can help teachers and professors identify whether student homework submissions, like essays or research papers, are AI-generated. This maintains academic integrity and ensures students engage in original thinking and writing. (ii) Copyright Protection~\cite{li2023protecting}: Watermarking can help detect if humans or AI write the given article. It protects the LLM owners' copyrights because the article publishers could make profits from the AI-generated content without proper acknowledgment.
(iii) Misinformation Monitor~\cite{megias2022architecture}: Watermarking can be used by social media platforms to detect and label watermarked AI-generated content automatically. It helps to combat the potential spread of misinformation or inauthentic content.

Adding post-hoc watermarks in LLM-generated texts can be methodologically categorized into~\cite{tang2023science}: (1) Rule-based watermarking, (2) Inference-time Watermarking, and (3) Neural-based Watermarking. 

\noindent\textbf{Rule-based  Watermarking}  This approach integrates watermarks into LLM-generated texts by manipulating linguistic features~\cite{he2022cater,yoo2023robust}, altering lexical properties~\cite{he2022protecting}, and substituting synonyms~\cite{munyer2023deeptextmark,yang2023watermarking}.
The rule-based watermarking approach aims to insert the synonym replacement or syntactic transformations as watermarks while ensuring the overall semantics are not distorted. 

\noindent\textbf{Inference-time Watermarking} Inference-time watermarking~\cite{kirchenbauer2023watermark,kirchenbauer2023reliability} approach inserts signatures at the LLM decoding stage. This approach divides the vocabulary into red/green lists and only allows LLM to decode tokens from the green list.  Some follow-up works~\cite{christ2023undetectable,zhao2023provable,huo2024token,liu2024adaptive} proposed different red/green list splitting algorithms or sampling algorithms from the green list probabilistic distribution to enhance the explainability, semantic preservation, and robustness of the inference-time watermarking. 

\noindent\textbf{Neural-based Watermarking} The neural-based approach~\cite{abdelnabi21oakland,zhang2023remark} encodes LLM-generated texts and associated message signatures through an end-to-end machine learning paradigm. It leverages a data-hiding network to infuse the watermark signatures into the LLM-generated texts and a data-revealing network to decode the signature from the watermark text. 
This facilitates the signatures' integration into the feature space of the watermarked text without compromising semantic fidelity.

\noindent\textbf{Comparison} The comparison of different watermarking methods under different evaluation criteria is in Table~\ref{tab:compare_scheme}.
The rule-based watermarking, like \baserule~\cite{he2022cater}, demonstrates effectiveness and efficiency by inserting watermarks in the linguistic attributes of the texts. However, the adversary may exploit syntactic transformations or synonym replacements to remove manually designed watermarking signatures from \baserule~\cite{he2022cater}.

The inference-time watermarking achieves resilience by embedding watermarks during each token decoding. However, \baseinfer~\cite{kirchenbauer2023watermark} introduces semantic discrepancies between the watermarked and original texts, undermining the LLM's fidelity. While \baseinferrb~\cite{kuditipudi2023robust} tries to improve the semantic preservation, the efficiency is compromised compared to \baseinfer~\cite{kirchenbauer2023watermark}.

The neural-based approach like REMARK-LLM~\cite{zhang2023remark} and \baseend~\cite{abdelnabi21oakland} leverages machine learning algorithms to embed watermarks into the LLM-generated texts without tampering with textual semantics. It achieves efficiency and robustness by embedding the watermarks through text feature space via lightweight language models. 

\begin{table}[!ht]
\centering
 \resizebox{\columnwidth}{!}{
\begin{tabular}{cccccc}
\toprule
Method & Effectiveness & Fidelity & Efficiency & Robustness  & Undetectability \\ \midrule
Rule-based~\cite{he2022cater,yoo2023robust} & \cmark & \cmark & \cmark & \xmark  & \cmark \\  
Inference-time~\cite{kirchenbauer2023watermark,kuditipudi2023robust} & \cmark & \xmark & \xmark &  \cmark  & \cmark\\  
Neural-based~\cite{zhang2023remark,abdelnabi21oakland} &  \cmark & \cmark &  \cmark  &   \cmark  & \cmark \\  
\bottomrule
\end{tabular}
}
\caption{Comparison of LLM-generated text watermarking schemes.}\label{tab:compare_scheme}
\end{table}

\subsection{Domain-Specific Data Watermarking}

Watermarking domain-specific data presents unique challenges compared to watermarking natural language content. The primary difficulty lies in maintaining the integrity and functionality of the specialized data while embedding watermarks. Unlike general text, where minor alterations may not significantly impact meaning, domain-specific data often requires precise formatting, syntax, or content to remain valid and useful. For example, watermarking codes~\cite{li2023protecting,lee2023wrote,yang2023towards} must ensure that the functionality and executability of the program are preserved, while watermarking medical data~\cite{anand2021watermarking,guo2009region,boussif2018secured} requires the retention of accurate terminology, proper nouns, and critical health information. These requirements add layers of complexity to the watermarking process, as any introduced changes must be carefully designed to avoid compromising the data's primary purpose or introducing errors.

\noindent\textbf{LLM-generated Code Watermarking} 
Rule-based transformation~\cite{li2023protecting,sun2023codemark,ning2024mcgmark} methods involve applying predefined rules to modify code structure or syntax without altering functionality, such as changing variable naming conventions, restructuring control flow, or inserting benign code segments. Neural network-based approaches~\cite{yang2023towards,yang2024srcmarker}, on the other hand, leverage deep learning models to learn transformation patterns that can embed watermarks while preserving code semantics. These models can be trained on large code repositories to understand the nuances of different programming languages and coding styles, enabling more sophisticated and adaptable watermarking strategies.

\noindent\textbf{LLM-generated Medical Data Watermarking} Medical data watermarking~\cite{kong2024protecting,guo2023proteinchat,huo2024multi} requires a delicate balance between data protection and maintaining the accuracy of critical health information. One key challenge is the correct handling of proper nouns, such as medical terms, drug names, and anatomical references, which must remain unaltered to preserve the data's integrity and correctness. 

\noindent\textbf{LLM-assisted Chip Design Watermarking} Chip design watermarking~\cite{rathor2023hard,tauhid2023survey,zhang2024automated,zhang2024icmarks} encodes unique and confidential signatures into the IC layouts to assist product owners in proof of ownership.
Design houses leverage this technique to enhance IP rights protection in the chip supply chain and detect unauthorized usages or replications. 
These watermarks are typically inserted in the logic design or physical design level. By carefully designing the insertion methodology, the watermark is encoded without compromising the IC layout quality.

\subsection{Potential Threats}

The adversary, as an end-user of the LLM cloud service, where has black access to the API. However, he does not have access to the trained watermarking models and the original LLM-generated outputs. 
The adversary attempts to exploit the LLM-generated content for malicious usage without being traced. Therefore, he performs attacks to detect and remove the signatures within the watermarked contents without distorting their semantics. 

The potential attacks to the watermarked content include~\cite{zhang2023remark,huo2024token}:

\begin{itemize}
\item \textit{\textbf{Attack 1 Watermark Detection Attack}}: The adversary uses statistical analysis or machine learning models to detect whether the texts are watermarked. 

\item \textit{\textbf{Attack 2 Text Edit Attack}}: The adversary doesn't have prior linguistic knowledge. By randomly deleting, adding, or substituting words within the content, he attempts to destroy the watermark while preserving the overall meanings.

\item \textit{\textbf{Attack 3 Text Rephrase Attack}}: The adversary can exploit open-source NLP models, such as T5, to remove watermarks. By feeding the content into these models, the adversary aims to generate a rephrased version of the original texts to remove the watermark.

\item \textit{\textbf{Attack 4 Re-watermarking Attack}}: The adversary dispatches the watermarked texts into another LLM watermarking framework like \sys{} and re-watermark it to remove the inserted signatures. 
 \end{itemize}

\section{Watermarking Acceleration}
In this section, we introduce software/hardware co-design methods~\cite{zhang2023systemization} to accelerate the previously mentioned watermark insertion and verification process. 

\subsection{Model Watermarking Acceleration}
For the LLM model watermarking, software/hardware co-design can accelerate the watermark insertion/verification process. To facilitate efficient hardware acceleration of watermark insertion and extraction processes, it is crucial to design algorithms with hardware-friendly characteristics. For example, the algorithms shall be restructured to exploit data-level parallelism~\cite{brakel2024model,stojkovic2024dynamollm,yu2024twinpilots}, allowing multiple watermark operations to be performed simultaneously on different portions of the model or input data. By carefully considering the target hardware architecture during the algorithm design phase, we can ensure that the watermarking processes are amenable to acceleration on platforms such as FPGAs, ASICs, or specialized AI accelerators. This co-design approach enhances the performance of watermark insertion and extraction and promotes seamless integration with existing hardware-accelerated deep-learning pipelines.

Hardware acceleration can significantly boost the speed of watermark insertion and extraction processes through various specialized architectures and optimizations. FPGAs, for example, can be configured to implement custom datapaths tailored specifically for watermarking operations~\cite{cao2020digital,lim2003fpga,hussain2022faststamp,weng2023adapting}, leveraging their reconfigurable nature to achieve high throughput and energy efficiency. These custom datapaths can include parallel processing elements for concurrent watermark computations, dedicated memory structures for efficient data access, and optimized arithmetic units for watermark-specific calculations. By leveraging these hardware acceleration techniques, the overhead of watermarking can be minimized, enabling the widespread adoption of model protection mechanisms without compromising the overall system performance.

\subsection{Content Watermarking Acceleration}
Content watermarking techniques, including neural-based and rule-based approaches, can benefit from acceleration strategies similar to those employed for model watermarking. Neural-based content watermarking, which often involves embedding watermarks into the content using neural networks, can leverage the same hardware acceleration techniques used for general deep learning inference. This includes utilizing specialized AI accelerators, FPGAs, or ASICs to parallelize the watermark embedding process across multiple data samples or content segments. Rule-based watermarking, which relies on predefined algorithms to insert watermarks, can be optimized for hardware acceleration by designing rules that map efficiently to parallel processing architectures.

Inference-based watermarking, however, involves encoding watermarks during the LLM inference stage. As such, the inference optimization techniques can benefit such watermark insertion procedures. Efficient decoding approaches, such as speculative decoding~\cite{leviathan2023fast,kim2024speculative,chen2023accelerating}, skip layer~\cite{fan2024not,huang2024raee,raposo2024mixture}, can be integrated with watermarking techniques to maintain high-quality output while inserting watermarks more efficiently. These approaches can dynamically adjust the computational resources allocated to watermarking based on the content complexity or watermark requirements, ensuring an optimal balance between watermarking effectiveness and generation speed.


\section{Future Direction}
In this section, we present a few future directions that will help enhance the current watermarking algorithm's robustness, applicability, and scalability. 

\subsection{Robustness against Evolving Attacks} 

As adversaries become more adept at evading current watermarking methods, future solutions must demonstrate enhanced robustness against text modifications, paraphrasing, and content modification attempts. Moreover, the watermarking schemes should offer improved traceability, enabling the distinction between multiple users and use cases, thereby facilitating more granular content attribution and ownership verification. 
Furthermore, as AI models increasingly generate and process multimodal data~\cite{liu2023llava,wu2023next,hu2024bliva,jia2023physical}, the watermarking techniques will need to address the challenges of embedding and detecting watermarks across diverse data types such as text, images, audio, and video. This will require the development of unified watermarking frameworks capable of seamlessly integrating and tracing information across multiple modalities, ensuring comprehensive protection and attribution in complex, multi-format AI-generated content.

\subsection{Adaptive Regulation and Authentication}  

As watermarking techniques become more sophisticated, there will be a need to establish clear guidelines and standards for their use~\cite{liang2024monitoring,longpre2024safe}, particularly in scenarios where watermarked content may be used as evidence in legal proceedings or for attribution in creative works. This could lead to the development of standardized watermarking protocols and verification mechanisms that are recognized across industries and jurisdictions. Additionally, adaptive watermarking, where the strength or presence of watermarks is dynamically adjusted based on the intended use of the generated content, balancing the rights of model owners with the needs of end-users.

\subsection{Explainability} 

Apart from identification or tracing, designing of the watermarking algorithms with explainability~\cite{rosales2024trustworthiness,shao2024explanation} in mind will benefit model users. By leveraging the model's inherent understanding and generation processes, model owners can embed richer watermarking information, such as version identifiers and user IDs, into both the LLM itself and its generated content. By aligning watermarking techniques with explainable AI methods, owners can create more sophisticated and informative watermarks that identify the source and provide insights into the watermarking framework's verification decision-making process.

\section{Conclusion}
In conclusion, this paper has provided a comprehensive overview of the challenges and opportunities in watermarking large language models (LLMs) and their generated content. We have explored a range of watermarking techniques for LLMs under various threat models and scenarios, as well as methods for watermarking LLM-generated content. By surveying the current watermarking frameworks and outlining future research directions, this paper contributes to the ongoing efforts to ensure the responsible use and protection of generative AI tools. As LLMs continue to evolve and shape the future, our works will contribute to the development of robust, efficient, and adaptable watermarking techniques that safeguard intellectual property rights, trace content origins, and mitigate the spread of AI-generated misinformation.

\section{Acknowledgments}
This work was supported by NSF TILOS AI Institute award number 2112665.

\bibliographystyle{IEEEtran}
\bibliography{bib}

\begin{thebibliography}{100}
\providecommand{\url}[1]{#1}
\csname url@samestyle\endcsname
\providecommand{\newblock}{\relax}
\providecommand{\bibinfo}[2]{#2}
\providecommand{\BIBentrySTDinterwordspacing}{\spaceskip=0pt\relax}
\providecommand{\BIBentryALTinterwordstretchfactor}{4}
\providecommand{\BIBentryALTinterwordspacing}{\spaceskip=\fontdimen2\font plus
\BIBentryALTinterwordstretchfactor\fontdimen3\font minus \fontdimen4\font\relax}
\providecommand{\BIBforeignlanguage}[2]{{%
\expandafter\ifx\csname l@#1\endcsname\relax
\typeout{** WARNING: IEEEtran.bst: No hyphenation pattern has been}%
\typeout{** loaded for the language `#1'. Using the pattern for}%
\typeout{** the default language instead.}%
\else
\language=\csname l@#1\endcsname
\fi
#2}}
\providecommand{\BIBdecl}{\relax}
\BIBdecl

\bibitem{touvron2023llama}
H.~Touvron, L.~Martin, K.~Stone, P.~Albert, A.~Almahairi, Y.~Babaei, N.~Bashlykov, S.~Batra, P.~Bhargava, S.~Bhosale \emph{et~al.}, ``Llama 2: Open foundation and fine-tuned chat models,'' \emph{arXiv preprint arXiv:2307.09288}, 2023.

\bibitem{abdin2024phi}
M.~Abdin, S.~A. Jacobs, A.~A. Awan, J.~Aneja, A.~Awadallah, H.~Awadalla, N.~Bach, A.~Bahree, A.~Bakhtiari, H.~Behl \emph{et~al.}, ``Phi-3 technical report: A highly capable language model locally on your phone,'' \emph{arXiv preprint arXiv:2404.14219}, 2024.

\bibitem{zhou2024survey}
P.~Zhou, L.~Wang, Z.~Liu, Y.~Hao, P.~Hui, S.~Tarkoma, and J.~Kangasharju, ``A survey on generative ai and llm for video generation, understanding, and streaming,'' \emph{arXiv preprint arXiv:2404.16038}, 2024.

\bibitem{shen2024data}
L.~Shen, H.~Li, Y.~Wang, and H.~Qu, ``From data to story: Towards automatic animated data video creation with llm-based multi-agent systems,'' \emph{arXiv preprint arXiv:2408.03876}, 2024.

\bibitem{roziere2023code}
B.~Roziere, J.~Gehring, F.~Gloeckle, S.~Sootla, I.~Gat, X.~E. Tan, Y.~Adi, J.~Liu, T.~Remez, J.~Rapin \emph{et~al.}, ``Code llama: Open foundation models for code,'' \emph{arXiv preprint arXiv:2308.12950}, 2023.

\bibitem{luo2023wizardcoder}
Z.~Luo, C.~Xu, P.~Zhao, Q.~Sun, X.~Geng, W.~Hu, C.~Tao, J.~Ma, Q.~Lin, and D.~Jiang, ``Wizardcoder: Empowering code large language models with evol-instruct,'' \emph{arXiv preprint arXiv:2306.08568}, 2023.

\bibitem{nakkab2024rome}
A.~Nakkab, S.~Q. Zhang, R.~Karri, and S.~Garg, ``Rome was not built in a single step: Hierarchical prompting for llm-based chip design,'' in \emph{Proceedings of the 2024 ACM/IEEE International Symposium on Machine Learning for CAD}, 2024, pp. 1--11.

\bibitem{liu2023chipnemo}
M.~Liu, T.-D. Ene, R.~Kirby, C.~Cheng, N.~Pinckney, R.~Liang, J.~Alben, H.~Anand, S.~Banerjee, I.~Bayraktaroglu \emph{et~al.}, ``Chipnemo: Domain-adapted llms for chip design,'' \emph{arXiv preprint arXiv:2311.00176}, 2023.

\bibitem{zhang2024automated}
R.~Zhang, R.~S. Rajarathnam, D.~Z. Pan, and F.~Koushanfar, ``Automated physical design watermarking leveraging graph neural networks,'' in \emph{Proceedings of the 2024 ACM/IEEE International Symposium on Machine Learning for CAD}, 2024, pp. 1--10.

\bibitem{wang2024jmlr}
J.~Wang, Z.~Yang, Z.~Yao, and H.~Yu, ``Jmlr: Joint medical llm and retrieval training for enhancing reasoning and professional question answering capability,'' \emph{arXiv preprint arXiv:2402.17887}, 2024.

\bibitem{panagoulias2024evaluating}
D.~P. Panagoulias, M.~Virvou, and G.~A. Tsihrintzis, ``Evaluating llm--generated multimodal diagnosis from medical images and symptom analysis,'' \emph{arXiv preprint arXiv:2402.01730}, 2024.

\bibitem{yuan2024continued}
D.~Yuan, E.~Rastogi, G.~Naik, J.~Chintagunta, S.~P. Rajagopal, F.~Zhao, S.~Goyal, and J.~Ward, ``A continued pretrained llm approach for automatic medical note generation,'' \emph{arXiv preprint arXiv:2403.09057}, 2024.

\bibitem{lin2024llm}
F.~Lin, D.~J. Kim \emph{et~al.}, ``When llm-based code generation meets the software development process,'' \emph{arXiv preprint arXiv:2403.15852}, 2024.

\bibitem{xia2024agentless}
C.~S. Xia, Y.~Deng, S.~Dunn, and L.~Zhang, ``Agentless: Demystifying llm-based software engineering agents,'' \emph{arXiv preprint arXiv:2407.01489}, 2024.

\bibitem{schulman2022chatgpt}
J.~Schulman, B.~Zoph, C.~Kim, J.~Hilton, J.~Menick, J.~Weng, J.~F.~C. Uribe, L.~Fedus, L.~Metz, M.~Pokorny \emph{et~al.}, ``Chatgpt: Optimizing language models for dialogue,'' \emph{OpenAI blog}, 2022.

\bibitem{OpenOrca}
W.~Lian, B.~Goodson, E.~Pentland, A.~Cook, C.~Vong, and "Teknium", ``Openorca: An open dataset of gpt augmented flan reasoning traces,'' \url{https://https://huggingface.co/Open-Orca/OpenOrca},, 2023.

\bibitem{lin2024qserve}
Y.~Lin, H.~Tang, S.~Yang, Z.~Zhang, G.~Xiao, C.~Gan, and S.~Han, ``Qserve: W4a8kv4 quantization and system co-design for efficient llm serving,'' \emph{arXiv preprint arXiv:2405.04532}, 2024.

\bibitem{lin2024awq}
J.~Lin, J.~Tang, H.~Tang, S.~Yang, W.-M. Chen, W.-C. Wang, G.~Xiao, X.~Dang, C.~Gan, and S.~Han, ``Awq: Activation-aware weight quantization for on-device llm compression and acceleration,'' \emph{Proceedings of Machine Learning and Systems}, vol.~6, pp. 87--100, 2024.

\bibitem{sheybani2023zkrownn}
N.~Sheybani, Z.~Ghodsi, R.~Kapila, and F.~Koushanfar, ``Zkrownn: Zero knowledge right of ownership for neural networks,'' in \emph{2023 60th ACM/IEEE Design Automation Conference (DAC)}.\hskip 1em plus 0.5em minus 0.4em\relax IEEE, 2023, pp. 1--6.

\bibitem{chen2019deepattest}
H.~Chen, C.~Fu, B.~D. Rouhani, J.~Zhao, and F.~Koushanfar, ``Deepattest: an end-to-end attestation framework for deep neural networks,'' in \emph{2019 ACM/IEEE 46th Annual International Symposium on Computer Architecture (ISCA)}.\hskip 1em plus 0.5em minus 0.4em\relax IEEE, 2019, pp. 487--498.

\bibitem{zhang2024emmark}
R.~Zhang and F.~Koushanfar, ``Emmark: Robust watermarks for ip protection of embedded quantized large language models,'' \emph{arXiv preprint arXiv:2402.17938}, 2024.

\bibitem{aldahmani2023cyber}
A.~Aldahmani, B.~Ouni, T.~Lestable, and M.~Debbah, ``Cyber-security of embedded iots in smart homes: challenges, requirements, countermeasures, and trends,'' \emph{IEEE Open Journal of Vehicular Technology}, vol.~4, pp. 281--292, 2023.

\bibitem{caviglione2023emerging}
L.~Caviglione, C.~Comito, M.~Guarascio, and G.~Manco, ``Emerging challenges and perspectives in deep learning model security: A brief survey,'' \emph{Systems and Soft Computing}, vol.~5, p. 200050, 2023.

\bibitem{kirchenbauer2023watermark}
J.~Kirchenbauer, J.~Geiping, Y.~Wen, J.~Katz, I.~Miers, and T.~Goldstein, ``A watermark for large language models,'' in \emph{International Conference on Machine Learning}.\hskip 1em plus 0.5em minus 0.4em\relax PMLR, 2023, pp. 17\,061--17\,084.

\bibitem{zhang2023remark}
R.~Zhang, S.~S. Hussain, P.~Neekhara, and F.~Koushanfar, ``Remark-llm: A robust and efficient watermarking framework for generative large language models,'' \emph{arXiv preprint arXiv:2310.12362}, 2023.

\bibitem{huo2024token}
M.~Huo, S.~A. Somayajula, Y.~Liang, R.~Zhang, F.~Koushanfar, and P.~Xie, ``Token-specific watermarking with enhanced detectability and semantic coherence for large language models,'' \emph{arXiv preprint arXiv:2402.18059}, 2024.

\bibitem{chen2023can}
C.~Chen and K.~Shu, ``Can llm-generated misinformation be detected?'' \emph{arXiv preprint arXiv:2309.13788}, 2023.

\bibitem{chen2023combating}
------, ``Combating misinformation in the age of llms: Opportunities and challenges,'' \emph{AI Magazine}, 2023.

\bibitem{vp2024llm}
S.~E. VP, R.~Dheepthi \emph{et~al.}, ``Llm-enhanced deepfake detection: Dense cnn and multi-modal fusion framework for precise multimedia authentication,'' in \emph{2024 International Conference on Advances in Data Engineering and Intelligent Computing Systems (ADICS)}.\hskip 1em plus 0.5em minus 0.4em\relax IEEE, 2024, pp. 1--6.

\bibitem{cai2023av}
Z.~Cai, S.~Ghosh, A.~P. Adatia, M.~Hayat, A.~Dhall, and K.~Stefanov, ``Av-deepfake1m: A large-scale llm-driven audio-visual deepfake dataset,'' \emph{arXiv preprint arXiv:2311.15308}, 2023.

\bibitem{jiang2024circuitnet}
X.~Jiang, Y.~Zhao, Y.~Lin, R.~Wang, R.~Huang \emph{et~al.}, ``Circuitnet 2.0: An advanced dataset for promoting machine learning innovations in realistic chip design environment,'' in \emph{The Twelfth International Conference on Learning Representations}, 2024.

\bibitem{zeng2020meddialog}
G.~Zeng, W.~Yang, Z.~Ju, Y.~Yang, S.~Wang, R.~Zhang, M.~Zhou, J.~Zeng, X.~Dong, R.~Zhang \emph{et~al.}, ``Meddialog: Large-scale medical dialogue datasets,'' in \emph{Proceedings of the 2020 conference on empirical methods in natural language processing (EMNLP)}, 2020, pp. 9241--9250.

\bibitem{liu2023reason}
Z.~Liu, H.~Hu, S.~Zhang, H.~Guo, S.~Ke, B.~Liu, and Z.~Wang, ``Reason for future, act for now: A principled framework for autonomous llm agents with provable sample efficiency,'' \emph{arXiv preprint arXiv:2309.17382}, 2023.

\bibitem{nam2024using}
D.~Nam, A.~Macvean, V.~Hellendoorn, B.~Vasilescu, and B.~Myers, ``Using an llm to help with code understanding,'' in \emph{Proceedings of the IEEE/ACM 46th International Conference on Software Engineering}, 2024, pp. 1--13.

\bibitem{zhang2023systemization}
R.~Zhang, S.~Hussain, H.~Chen, M.~Javaheripi, and F.~Koushanfar, ``Systemization of knowledge: Robust deep learning using hardware-software co-design in centralized and federated settings,'' \emph{ACM Transactions on Design Automation of Electronic Systems}, vol.~28, no.~6, pp. 1--32, 2023.

\bibitem{wang2024towards}
Y.~Wang, Y.~Chen, Z.~Li, Z.~Tang, R.~Guo, X.~Wang, Q.~Wang, A.~C. Zhou, and X.~Chu, ``Towards efficient and reliable llm serving: A real-world workload study,'' \emph{arXiv preprint arXiv:2401.17644}, 2024.

\bibitem{li2024llm}
B.~Li, Y.~Jiang, V.~Gadepally, and D.~Tiwari, ``Llm inference serving: Survey of recent advances and opportunities,'' \emph{arXiv preprint arXiv:2407.12391}, 2024.

\bibitem{zeng2024flightllm}
S.~Zeng, J.~Liu, G.~Dai, X.~Yang, T.~Fu, H.~Wang, W.~Ma, H.~Sun, S.~Li, Z.~Huang \emph{et~al.}, ``Flightllm: Efficient large language model inference with a complete mapping flow on fpgas,'' in \emph{Proceedings of the 2024 ACM/SIGDA International Symposium on Field Programmable Gate Arrays}, 2024, pp. 223--234.

\bibitem{huang2024edgellm}
M.~Huang, A.~Shen, K.~Li, H.~Peng, B.~Li, and H.~Yu, ``Edgellm: A highly efficient cpu-fpga heterogeneous edge accelerator for large language models,'' \emph{arXiv preprint arXiv:2407.21325}, 2024.

\bibitem{lai2024lcm}
C.~Lai, Z.~Zhou, A.~Poptani, and W.~Zhang, ``Lcm: Llm-focused hybrid spm-cache architecture with cache management for multi-core ai accelerators,'' in \emph{Proceedings of the 38th ACM International Conference on Supercomputing}, 2024, pp. 62--73.

\bibitem{li2024double}
S.~Li, L.~Yao, J.~Gao, L.~Zhang, and Y.~Li, ``Double-i watermark: Protecting model copyright for llm fine-tuning,'' \emph{arXiv preprint arXiv:2402.14883}, 2024.

\bibitem{wei2024proving}
J.~T.-Z. Wei, R.~Y. Wang, and R.~Jia, ``Proving membership in llm pretraining data via data watermarks,'' \emph{arXiv preprint arXiv:2402.10892}, 2024.

\bibitem{li2023watermarking}
L.~Li, B.~Jiang, P.~Wang, K.~Ren, H.~Yan, and X.~Qiu, ``Watermarking llms with weight quantization,'' \emph{arXiv preprint arXiv:2310.11237}, 2023.

\bibitem{li2023plmmark}
P.~Li, P.~Cheng, F.~Li, W.~Du, H.~Zhao, and G.~Liu, ``Plmmark: a secure and robust black-box watermarking framework for pre-trained language models,'' in \emph{Proceedings of the AAAI Conference on Artificial Intelligence}, vol.~37, no.~12, 2023, pp. 14\,991--14\,999.

\bibitem{gu2022watermarking}
C.~Gu, C.~Huang, X.~Zheng, K.-W. Chang, and C.-J. Hsieh, ``Watermarking pre-trained language models with backdooring,'' \emph{arXiv preprint arXiv:2210.07543}, 2022.

\bibitem{xu2024instructional}
J.~Xu, F.~Wang, M.~D. Ma, P.~W. Koh, C.~Xiao, and M.~Chen, ``Instructional fingerprinting of large language models,'' \emph{arXiv preprint arXiv:2401.12255}, 2024.

\bibitem{chen2020specmark}
H.~Chen, B.~D. Rouhani, and F.~Koushanfar, ``Specmark: A spectral watermarking framework for ip protection of speech recognition systems.'' in \emph{Interspeech}, 2020, pp. 2312--2316.

\bibitem{wu2024secgpt}
Y.~Wu, F.~Roesner, T.~Kohno, N.~Zhang, and U.~Iqbal, ``Secgpt: An execution isolation architecture for llm-based systems,'' \emph{arXiv preprint arXiv:2403.04960}, 2024.

\bibitem{yang2024first}
H.~Yang, D.~Zhang, Y.~Zhao, Y.~Li, and Y.~Liu, ``A first look at efficient and secure on-device llm inference against kv leakage,'' \emph{arXiv preprint arXiv:2409.04040}, 2024.

\bibitem{huang2024fast}
W.~Huang, Y.~Wang, A.~Cheng, A.~Zhou, C.~Yu, and L.~Wang, ``A fast, performant, secure distributed training framework for llm,'' in \emph{ICASSP 2024-2024 IEEE International Conference on Acoustics, Speech and Signal Processing (ICASSP)}.\hskip 1em plus 0.5em minus 0.4em\relax IEEE, 2024, pp. 4800--4804.

\bibitem{sabt2015trusted}
M.~Sabt, M.~Achemlal, and A.~Bouabdallah, ``Trusted execution environment: What it is, and what it is not,'' in \emph{2015 IEEE Trustcom/BigDataSE/Ispa}, vol.~1.\hskip 1em plus 0.5em minus 0.4em\relax IEEE, 2015, pp. 57--64.

\bibitem{jauernig2020trusted}
P.~Jauernig, A.-R. Sadeghi, and E.~Stapf, ``Trusted execution environments: properties, applications, and challenges,'' \emph{IEEE Security \& Privacy}, vol.~18, no.~2, pp. 56--60, 2020.

\bibitem{ekberg2013trusted}
J.-E. Ekberg, K.~Kostiainen, and N.~Asokan, ``Trusted execution environments on mobile devices,'' in \emph{Proceedings of the 2013 ACM SIGSAC conference on Computer \& communications security}, 2013, pp. 1497--1498.

\bibitem{boenisch2021systematic}
F.~Boenisch, ``A systematic review on model watermarking for neural networks,'' \emph{Frontiers in big Data}, vol.~4, p. 729663, 2021.

\bibitem{darvish2019deepsigns}
B.~Darvish~Rouhani, H.~Chen, and F.~Koushanfar, ``Deepsigns: An end-to-end watermarking framework for ownership protection of deep neural networks,'' in \emph{Proceedings of the twenty-fourth international conference on architectural support for programming languages and operating systems}, 2019, pp. 485--497.

\bibitem{lancaster2023artificial}
T.~Lancaster, ``Artificial intelligence, text generation tools and chatgpt--does digital watermarking offer a solution?'' \emph{International Journal for Educational Integrity}, vol.~19, no.~1, p.~10, 2023.

\bibitem{li2023protecting}
Z.~Li, C.~Wang, S.~Wang, and C.~Gao, ``Protecting intellectual property of large language model-based code generation apis via watermarks,'' in \emph{CCS}, 2023, pp. 2336--2350.

\bibitem{megias2022architecture}
D.~Meg{\'\i}as, M.~Kuribayashi, A.~Rosales, K.~Cabaj, and W.~Mazurczyk, ``Architecture of a fake news detection system combining digital watermarking, signal processing, and machine learning,'' \emph{Journal of Wireless Mobile Networks, Ubiquitous Computing, and Dependable Applications (JoWUA), 2022, 13 (1): 33-55,}, 2022.

\bibitem{tang2023science}
R.~Tang, Y.-N. Chuang, and X.~Hu, ``The science of detecting llm-generated texts,'' \emph{Communications of the ACM}, 2024.

\bibitem{he2022cater}
X.~He, Q.~Xu, Y.~Zeng, L.~Lyu, F.~Wu, J.~Li, and R.~Jia, ``Cater: Intellectual property protection on text generation apis via conditional watermarks,'' \emph{Advances in Neural Information Processing Systems}, vol.~35, pp. 5431--5445, 2022.

\bibitem{yoo2023robust}
K.~Yoo, W.~Ahn, J.~Jang, and N.~Kwak, ``Robust multi-bit natural language watermarking through invariant features,'' in \emph{Proceedings of the 61st Annual Meeting of the Association for Computational Linguistics (Volume 1: Long Papers)}, 2023, pp. 2092--2115.

\bibitem{he2022protecting}
X.~He, Q.~Xu, L.~Lyu, F.~Wu, and C.~Wang, ``Protecting intellectual property of language generation apis with lexical watermark,'' in \emph{Proceedings of the AAAI Conference on Artificial Intelligence}, vol.~36, no.~10, 2022, pp. 10\,758--10\,766.

\bibitem{munyer2023deeptextmark}
T.~Munyer and X.~Zhong, ``Deeptextmark: Deep learning based text watermarking for detection of large language model generated text,'' \emph{arXiv preprint arXiv:2305.05773}, 2023.

\bibitem{yang2023watermarking}
X.~Yang, K.~Chen, W.~Zhang, C.~Liu, Y.~Qi, J.~Zhang, H.~Fang, and N.~Yu, ``Watermarking text generated by black-box language models,'' \emph{arXiv preprint arXiv:2305.08883}, 2023.

\bibitem{kirchenbauer2023reliability}
J.~Kirchenbauer, J.~Geiping, Y.~Wen, M.~Shu, K.~Saifullah, K.~Kong, K.~Fernando, A.~Saha, M.~Goldblum, and T.~Goldstein, ``On the reliability of watermarks for large language models,'' \emph{arXiv preprint arXiv:2306.04634}, 2023.

\bibitem{christ2023undetectable}
M.~Christ, S.~Gunn, and O.~Zamir, ``Undetectable watermarks for language models,'' \emph{arXiv preprint arXiv:2306.09194}, 2023.

\bibitem{zhao2023provable}
X.~Zhao, P.~Ananth, L.~Li, and Y.-X. Wang, ``Provable robust watermarking for ai-generated text,'' \emph{arXiv preprint arXiv:2306.17439}, 2023.

\bibitem{liu2024adaptive}
Y.~Liu and Y.~Bu, ``Adaptive text watermark for large language models,'' \emph{arXiv preprint arXiv:2401.13927}, 2024.

\bibitem{abdelnabi21oakland}
S.~Abdelnabi and M.~Fritz, ``Adversarial watermarking transformer: Towards tracing text provenance with data hiding,'' in \emph{42nd IEEE Symposium on Security and Privacy}, 2021.

\bibitem{kuditipudi2023robust}
R.~Kuditipudi, J.~Thickstun, T.~Hashimoto, and P.~Liang, ``Robust distortion-free watermarks for language models,'' \emph{arXiv preprint arXiv:2307.15593}, 2023.

\bibitem{lee2023wrote}
T.~Lee, S.~Hong, J.~Ahn, I.~Hong, H.~Lee, S.~Yun, J.~Shin, and G.~Kim, ``Who wrote this code? watermarking for code generation,'' \emph{arXiv preprint arXiv:2305.15060}, 2023.

\bibitem{yang2023towards}
B.~Yang, W.~Li, L.~Xiang, and B.~Li, ``Towards code watermarking with dual-channel transformations,'' \emph{arXiv preprint arXiv:2309.00860}, 2023.

\bibitem{anand2021watermarking}
A.~Anand and A.~K. Singh, ``Watermarking techniques for medical data authentication: a survey,'' \emph{Multimedia Tools and Applications}, vol.~80, no.~20, pp. 30\,165--30\,197, 2021.

\bibitem{guo2009region}
X.~Guo and T.-g. Zhuang, ``A region-based lossless watermarking scheme for enhancing security of medical data,'' \emph{Journal of digital imaging}, vol.~22, pp. 53--64, 2009.

\bibitem{boussif2018secured}
M.~Boussif, N.~Aloui, and A.~Cherif, ``Secured cloud computing for medical data based on watermarking and encryption,'' \emph{IET Networks}, vol.~7, no.~5, pp. 294--298, 2018.

\bibitem{sun2023codemark}
Z.~Sun, X.~Du, F.~Song, and L.~Li, ``Codemark: Imperceptible watermarking for code datasets against neural code completion models,'' in \emph{Proceedings of the 31st ACM Joint European Software Engineering Conference and Symposium on the Foundations of Software Engineering}, 2023, pp. 1561--1572.

\bibitem{ning2024mcgmark}
K.~Ning, J.~Chen, Q.~Zhong, T.~Zhang, Y.~Wang, W.~Li, Y.~Zhang, W.~Zhang, and Z.~Zheng, ``Mcgmark: An encodable and robust online watermark for llm-generated malicious code,'' \emph{arXiv preprint arXiv:2408.01354}, 2024.

\bibitem{yang2024srcmarker}
B.~Yang, W.~Li, L.~Xiang, and B.~Li, ``Srcmarker: Dual-channel source code watermarking via scalable code transformations,'' in \emph{2024 IEEE Symposium on Security and Privacy (SP)}.\hskip 1em plus 0.5em minus 0.4em\relax IEEE Computer Society, 2024, pp. 97--97.

\bibitem{kong2024protecting}
C.~Kong, R.~Xu, W.~Chen, J.~Chen, and Z.~Yin, ``Protecting copyright of medical pre-trained language models: Training-free backdoor watermarking,'' \emph{arXiv preprint arXiv:2409.10570}, 2024.

\bibitem{guo2023proteinchat}
H.~Guo, M.~Huo, R.~Zhang, and P.~Xie, ``Proteinchat: Towards achieving chatgpt-like functionalities on protein 3d structures,'' \emph{Authorea Preprints}, 2023.

\bibitem{huo2024multi}
M.~Huo, H.~Guo, X.~Cheng, D.~Singh, H.~Rahmani, S.~Li, P.~Gerlof, T.~Ideker, D.~A. Grotjahn, E.~Villa \emph{et~al.}, ``Multi-modal large language model enables protein function prediction,'' \emph{bioRxiv}, pp. 2024--08, 2024.

\bibitem{rathor2023hard}
M.~Rathor and G.~P. Rathor, ``Hard-sign: A hardware watermarking scheme using dated handwritten signature,'' \emph{IEEE Design \& Test}, 2023.

\bibitem{tauhid2023survey}
A.~Tauhid, L.~Xu, M.~Rahman, and E.~Tomai, ``A survey on security analysis of machine learning-oriented hardware and software intellectual property,'' \emph{High-Confidence Computing}, p. 100114, 2023.

\bibitem{zhang2024icmarks}
R.~Zhang, R.~S. Rajarathnam, D.~Z. Pan, and F.~Koushanfar, ``Icmarks: A robust watermarking framework for integrated circuit physical design ip protection,'' \emph{arXiv preprint arXiv:2404.18407}, 2024.

\bibitem{brakel2024model}
F.~Brakel, U.~Odyurt, and A.-L. Varbanescu, ``Model parallelism on distributed infrastructure: A literature review from theory to llm case-studies,'' \emph{arXiv preprint arXiv:2403.03699}, 2024.

\bibitem{stojkovic2024dynamollm}
J.~Stojkovic, C.~Zhang, {\'I}.~Goiri, J.~Torrellas, and E.~Choukse, ``Dynamollm: Designing llm inference clusters for performance and energy efficiency,'' \emph{arXiv preprint arXiv:2408.00741}, 2024.

\bibitem{yu2024twinpilots}
C.~Yu, T.~Wang, Z.~Shao, L.~Zhu, X.~Zhou, and S.~Jiang, ``Twinpilots: A new computing paradigm for gpu-cpu parallel llm inference,'' in \emph{Proceedings of the 17th ACM International Systems and Storage Conference}, 2024, pp. 91--103.

\bibitem{cao2020digital}
Y.~Cao, F.~Yu, and Y.~Tang, ``A digital watermarking encryption technique based on fpga cloud accelerator,'' \emph{IEEE Access}, vol.~8, pp. 11\,800--11\,814, 2020.

\bibitem{lim2003fpga}
H.~Lim, S.-Y. Park, S.-J. Kang, and W.-H. Cho, ``Fpga implementation of image watermarking algorithm for a digital camera,'' in \emph{2003 IEEE Pacific Rim Conference on Communications Computers and Signal Processing (PACRIM 2003)(Cat. No. 03CH37490)}, vol.~2.\hskip 1em plus 0.5em minus 0.4em\relax IEEE, 2003, pp. 1000--1003.

\bibitem{hussain2022faststamp}
S.~Hussain, N.~Sheybani, P.~Neekhara, X.~Zhang, J.~Duarte, and F.~Koushanfar, ``Faststamp: Accelerating neural steganography and digital watermarking of images on fpgas,'' in \emph{Proceedings of the 41st IEEE/ACM International Conference on Computer-Aided Design}, 2022, pp. 1--9.

\bibitem{weng2023adapting}
O.~Weng, G.~Marcano, V.~Loncar, A.~Khodamoradi, N.~Sheybani, F.~Koushanfar, K.~Denolf, J.~M. Duarte, and R.~Kastner, ``Adapting skip connections for resource-efficient fpga inference,'' in \emph{Proceedings of the 2023 ACM/SIGDA International Symposium on Field Programmable Gate Arrays}, 2023, pp. 229--229.

\bibitem{leviathan2023fast}
Y.~Leviathan, M.~Kalman, and Y.~Matias, ``Fast inference from transformers via speculative decoding,'' in \emph{International Conference on Machine Learning}.\hskip 1em plus 0.5em minus 0.4em\relax PMLR, 2023, pp. 19\,274--19\,286.

\bibitem{kim2024speculative}
S.~Kim, K.~Mangalam, S.~Moon, J.~Malik, M.~W. Mahoney, A.~Gholami, and K.~Keutzer, ``Speculative decoding with big little decoder,'' \emph{Advances in Neural Information Processing Systems}, vol.~36, 2024.

\bibitem{chen2023accelerating}
C.~Chen, S.~Borgeaud, G.~Irving, J.-B. Lespiau, L.~Sifre, and J.~Jumper, ``Accelerating large language model decoding with speculative sampling,'' \emph{arXiv preprint arXiv:2302.01318}, 2023.

\bibitem{fan2024not}
S.~Fan, X.~Jiang, X.~Li, X.~Meng, P.~Han, S.~Shang, A.~Sun, Y.~Wang, and Z.~Wang, ``Not all layers of llms are necessary during inference,'' \emph{arXiv preprint arXiv:2403.02181}, 2024.

\bibitem{huang2024raee}
L.~Huang, S.~Wu, Y.~Cui, Y.~Xiong, X.~Liu, T.-W. Kuo, N.~Guan, and C.~J. Xue, ``Raee: A training-free retrieval-augmented early exiting framework for efficient inference,'' \emph{arXiv preprint arXiv:2405.15198}, 2024.

\bibitem{raposo2024mixture}
D.~Raposo, S.~Ritter, B.~Richards, T.~Lillicrap, P.~C. Humphreys, and A.~Santoro, ``Mixture-of-depths: Dynamically allocating compute in transformer-based language models,'' \emph{arXiv preprint arXiv:2404.02258}, 2024.

\bibitem{liu2023llava}
H.~Liu, C.~Li, Q.~Wu, and Y.~J. Lee, ``Visual instruction tuning,'' \emph{Advances in neural information processing systems}, vol.~36, 2024.

\bibitem{wu2023next}
S.~Wu, H.~Fei, L.~Qu, W.~Ji, and T.-S. Chua, ``Next-gpt: Any-to-any multimodal llm,'' \emph{arXiv preprint arXiv:2309.05519}, 2023.

\bibitem{hu2024bliva}
W.~Hu, Y.~Xu, Y.~Li, W.~Li, Z.~Chen, and Z.~Tu, ``Bliva: A simple multimodal llm for better handling of text-rich visual questions,'' in \emph{Proceedings of the AAAI Conference on Artificial Intelligence}, vol.~38, no.~3, 2024, pp. 2256--2264.

\bibitem{jia2023physical}
S.~Jia, S.~Wang, T.-M. Li, and A.~Chern, ``Physical cyclic animations,'' \emph{Proceedings of the ACM on Computer Graphics and Interactive Techniques}, vol.~6, no.~3, pp. 1--18, 2023.

\bibitem{liang2024monitoring}
W.~Liang, Z.~Izzo, Y.~Zhang, H.~Lepp, H.~Cao, X.~Zhao, L.~Chen, H.~Ye, S.~Liu, Z.~Huang \emph{et~al.}, ``Monitoring ai-modified content at scale: A case study on the impact of chatgpt on ai conference peer reviews,'' \emph{arXiv preprint arXiv:2403.07183}, 2024.

\bibitem{longpre2024safe}
S.~Longpre, S.~Kapoor, K.~Klyman, A.~Ramaswami, R.~Bommasani, B.~Blili-Hamelin, Y.~Huang, A.~Skowron, Z.-X. Yong, S.~Kotha \emph{et~al.}, ``A safe harbor for ai evaluation and red teaming,'' \emph{arXiv preprint arXiv:2403.04893}, 2024.

\bibitem{rosales2024trustworthiness}
A.~Rosales, A.~Malanowska, T.~Koohpayeh~Araghi, M.~Kuribayashi, M.~Kowalczyk, D.~Blanche-Tarrag{\'o}, W.~Mazurczyk, and D.~Meg{\'\i}as, ``Trustworthiness and explainability of a watermarking and machine learning-based system for image modification detection to combat disinformation,'' in \emph{Proceedings of the 19th International Conference on Availability, Reliability and Security}, 2024, pp. 1--10.

\bibitem{shao2024explanation}
S.~Shao, Y.~Li, H.~Yao, Y.~He, Z.~Qin, and K.~Ren, ``Explanation as a watermark: Towards harmless and multi-bit model ownership verification via watermarking feature attribution,'' \emph{arXiv preprint arXiv:2405.04825}, 2024.

\end{thebibliography}
 
\end{document}